\documentclass[aps, pre, floatfix,
twocolumn
]{revtex4-1}

\usepackage{graphicx}% Include figure files
\usepackage{bm}% bold math
\usepackage{subfigure}
\usepackage{hyperref}
\usepackage{amsmath}
\usepackage{color}

\begin{document}
%\preprint{APS/123-QED}

\title{Saturated random packing built of arbitrary polygons under random sequential adsorption protocol}

\author{Micha\l{} Cie\'sla}
  \email{michal.ciesla@uj.edu.pl}
\author{Piotr Kubala}
  \email{pkua.log@gmail.com}
\affiliation{M.\ Smoluchowski Institute of Physics, Department of Statistical Physics, Jagiellonian University, \L{}ojasiewicza 11, 30-348 Krak\'ow, Poland}
\author{Ge Zhang}
    \email{gezhang@alumni.princeton.edu}
\affiliation{Department of Physics, University of Pennsylvania, Philadelphia, Pennsylvania 19104, USA}

\date{\today}% It is always \today, today,

\begin{abstract}
Random packings and their properties are a popular and active field of research. Numerical algorithms that can efficiently generate them are useful tools in their study. This paper focuses on random packings produced according to the random sequential adsorption (RSA) protocol. Developing the idea presented in [G. Zhang, Phys. Rev. E {\bf 97}, 043311 (2018)], where saturated random packings built of regular polygons were studied, we create an algorithm that generates strictly saturated packings built of any polygons. Then, the algorithm was used to determine the packing fractions for arbitrary triangles. The highest mean packing density, $0.552814 \pm 0.000063$, was observed for triangles of side lengths $0.63:1:1$. Additionally, microstructural properties of such packings, kinetics of their growth as well as distributions of saturated packing fractions and the number of RSA iterations needed to reach saturation were analyzed.
\end{abstract}

%
%\pacs{02.70.Tt, 05.10.Ln, 68.43.Fg}
%
%\vspace{2pc} \noindent{\it Keywords}: random packings, random sequential adsorption, algorithm to generate saturated packings
\maketitle
\section{Introduction}
Random sequential adsorption (RSA) is a numerical protocol to generate random packings of particles with a given shape \cite{Evans1993, Talbot2000}. It is based on sequential trials to add a randomly placed particle to a packing. The particle is added when it does not intersect with any previously placed particle; otherwise, it is removed. When added, it will hold its position and orientation inside the packing forever. The process is continued until the packing is saturated, i.e., there is no area large enough to place any new particle.

The history of RSA began in 1939 when Flory was studying random attachment of atomic groups to a long polymer. In 1959, Renyi solved the so-called ``car parking problem'' \cite{Renyi1958}, which was a one-dimensional implementation of RSA. The popularity of RSA, however, owes to Feder, who noted that RSA packings resemble the structure of monolayers created during irreversible adsorption processes \cite{Feder1980}. RSA is also studied as one of the simplest yet not trivial protocols to generate random packings, which takes into account excluded volume effects. In contrast to random close packings, which are the most popular models of granular matter, saturated RSA packings have well-defined mean densities \cite{Torquato2000}. 

\begin{figure}[htb]
    \centering
    \subfigure[]{\includegraphics[width=0.22\columnwidth]{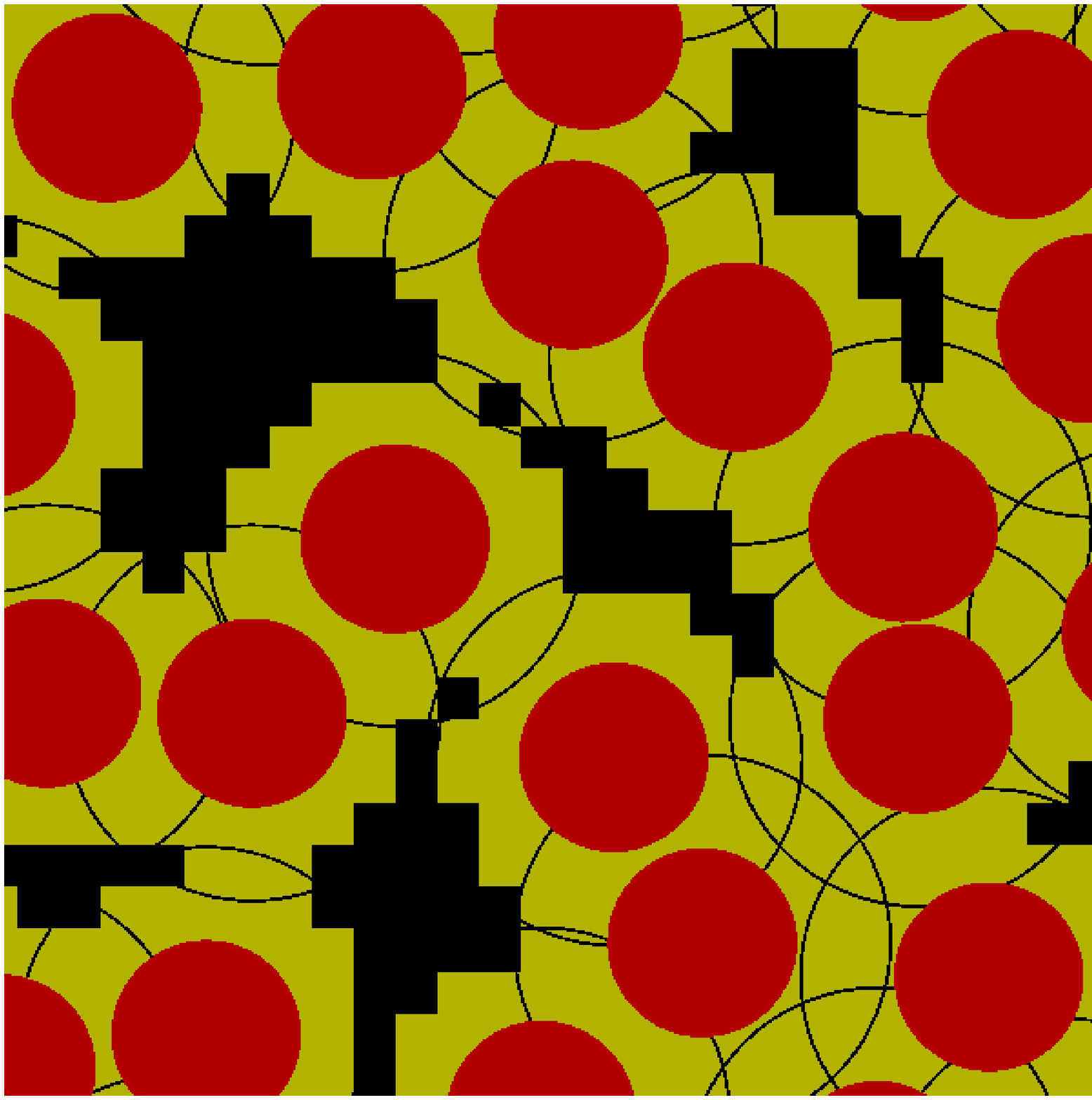}}
    \subfigure[]{\includegraphics[width=0.22\columnwidth]{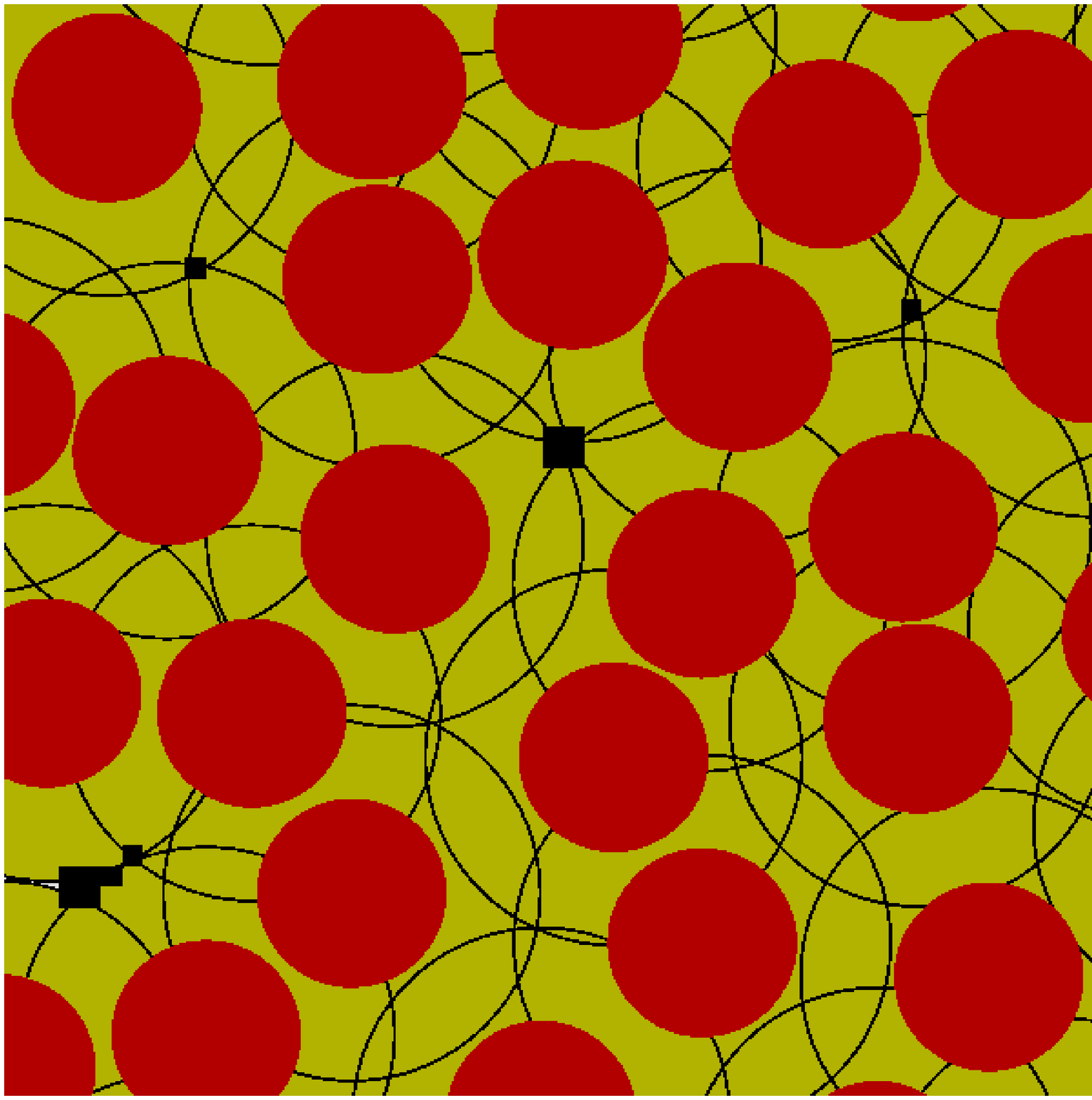}}
    \subfigure[]{\includegraphics[width=0.22\columnwidth]{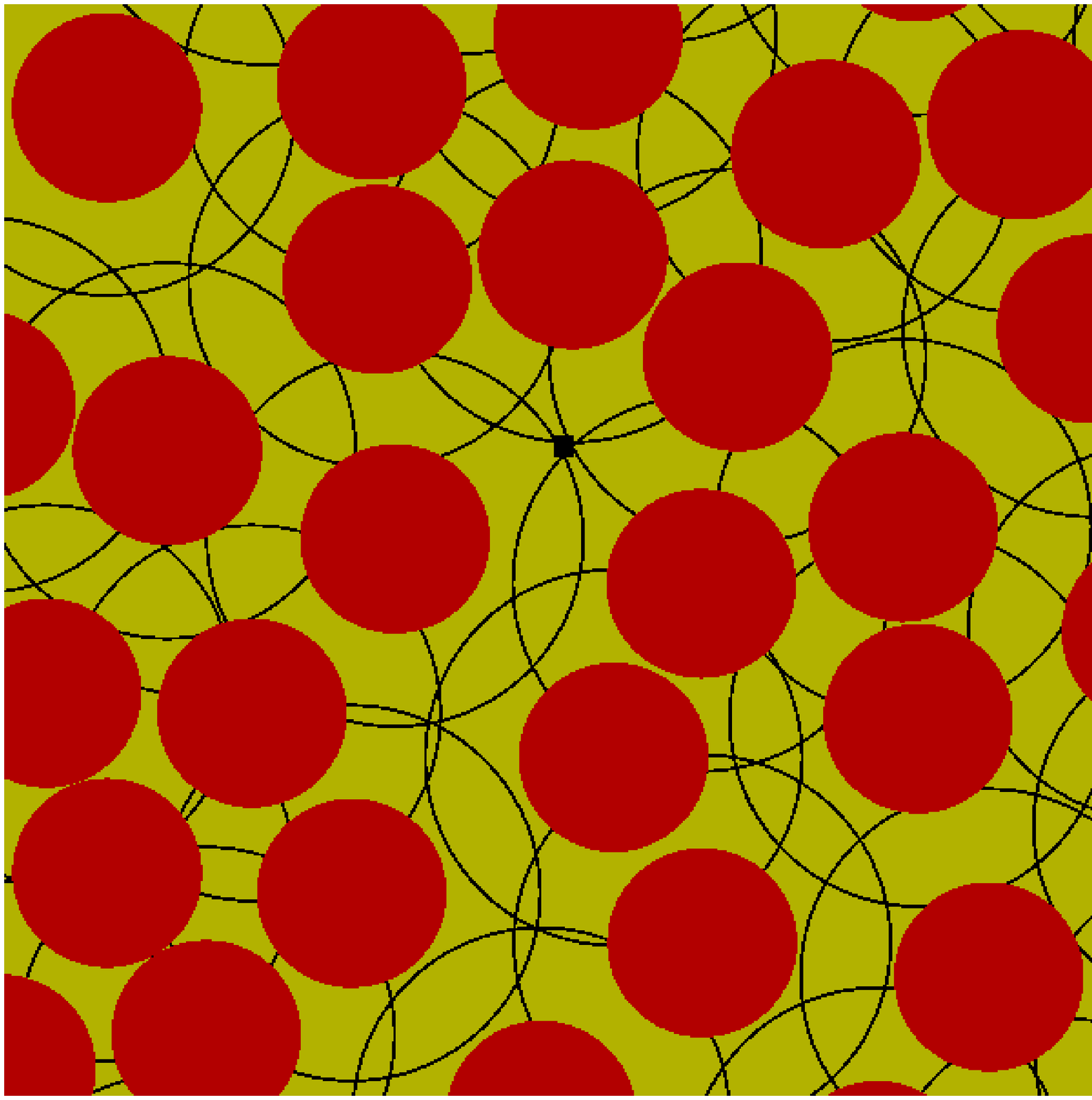}}
    \subfigure[]{\includegraphics[width=0.22\columnwidth]{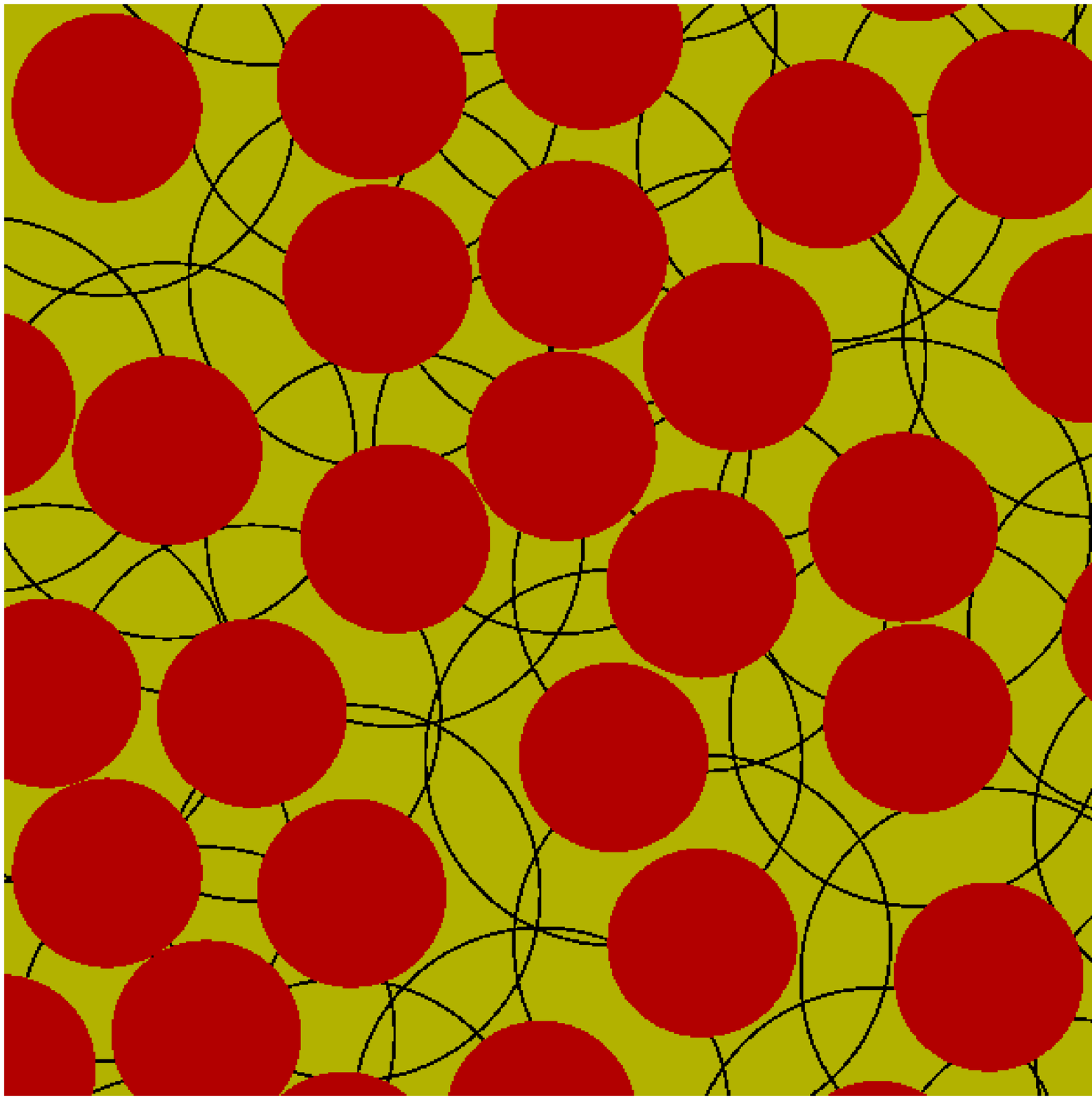}}
    \caption{(Color online) Stages involved in generating an RSA packing of 2D discs. Red discs represent non-penetrable particles. Yellow color indicates regions where it is impossible to place the center of another disc due to intersection with the existing ones. Black square voxels approximate an area where placing of a subsequent disc is possible. In each subsequent stage, these voxels are divided into $2^d$ sub-voxels to better approximate available space as well as are filled up with new discs. In panel (d), there are no black voxels; thus, the packing is saturated.}
    \label{fig:examples}
    \label{fig:disks}
\end{figure}

One of the biggest issues of the RSA protocol is its inefficiency for almost saturated packings. When the probability of successfully adding another particle without intersect to the packing is tiny, the algorithm needs a tremendous amount of random trials to insert any new particle. Moreover, even if the packing is already saturated, it is often difficult to ascertain saturation in order to stop adding subsequent particles to the packing. The most common way to deal with this problem is to stop the simulation after some large but arbitrarily defined number of iterations and then estimate interesting properties of saturated packings ({\it e.g.,} its density) by extrapolating packing growth kinetics \cite{Pomeau1980, Swendsen1981}. Although such an approach can give pretty accurate results \cite{Ciesla2016}, it requires large amount of computer resources and still is burdened with systematic error due to the estimation. The best method so far to deal with this problem is to trace regions where the next particle can be added. It allows one to significantly speed up packing generation because subsequent attempts to place an object can be restricted only to these regions. Moreover, when all such regions vanish, insertion attempts may cease since the packing is known to be saturated. The method is illustrated in Fig.\ref{fig:disks} for two-dimensional discs.

It has been successfully applied for packing built of spherically symmetric shapes \cite{Wang1994, Ebeida2012, Zhang2013} and, recently, for ellipses and spherocylinders \cite{Haiduk2018}, regular polygons \cite{Zhang2018}, rectangles \cite{Kasperek2018} and any shapes build of several disks \cite{Ciesla2020}.

The aim of this study is to show how the idea used for generating saturated random packing built of regular polygons \cite{Zhang2018} can be expanded to work for any polygon. It is worth nothing, that together with a recent algorithm presented in Ref.\cite{Ciesla2020}, this method allows one to generate strictly saturated RSA packings for a large range of two-dimensional shapes. Therefore it enables us to model effectively the majority of monolayers produced in irreversible adsorption experiments. For example, the presented algorithm is used here to determine the shape of triangles that build the densest RSA packing. Although the RSA packings built of triangles does not have a direct application in modelling adsorption monolayers of realistic molecules, their study may help answering some fundamental questions regarding random packing properties. Firstly, it was observed that shapes of higher symmetry, like disks or spheres, form less dense packings than anisotropic shapes like ellipses or ellipsoids \cite{Vigil1989,Viot1992,Donev2004,Man2005,Baule2013,Haiduk2018}. Triangles do not have continuous symmetry groups like disks or spheres, but they may have $D_3$ dihedral symmetry (equilateral triangles) or axial symmetry (isosceles tringles). It is important to check if these symmetries influence the properties of the obtained packings. It was observed that in the case of rectangles, the local minimum of the packing fraction is observed for squares \cite{Vigil1989,Kasperek2018}. Another recent study of random packings built of shapes with tetrahedral and octahedral symmetry shows that it is not true that octahedral symmetry will always lead to less dense packings than tetrahedral symmetry \cite{Kubala2019}. Secondly, recent analytical results obtained by Baule shows that the kinetics of packing growth near saturation limit may depend on properties of the contact function -- the touching distance between two objects as a function of their relative orientation \cite{Baule2017}. For example, the long-time scaling of the kinetics is different between packings built of ellipses and rectangles with centres placed on a one dimensional line. This has not been observed so far for two dimensional packings that are almost saturated \cite{Viot1992, Ciesla2014, Ciesla2016} but now, using the algorithm that generates strictly saturated packings, it is possible to check this arbitrarily close to saturation limit. Additionally it would be possible to check if the scaling of the number of RSA iterations needed to reach saturation is governed by the same exponent as the kinetics of packing growth \cite{Ciesla2017}.   

\section{Algorithm}
The crucial stage of the algorithm performance is when it determines if a given voxel can be eliminated due to impossibility to add another particle inside it. In case of discs or spherically symmetric objects, a voxel can be eliminated if the distance from each of its vertices to the center of any already added disc is smaller than $2r$, where $r$ is the disc's radius \cite{Wang1994,Zhang2013,Haiduk2018, Kasperek2018}. Because a voxel is convex, it is removed when all its corners are inside the excluded area around already placed particle. Another analytic approach is to define the intersection function and estimate its behavior inside a voxel. For example, in case of disc, such function can be defined as:
\begin{equation}
F_{disk}(x, y) = (x-x_0)^2 + (y-y_0)^2 - 4 r^2,
\label{Fdisk}
\end{equation}
where $(x_0, y_0)$ are coordinates of the disc center from the packing, and $r$ is its radius. Thus, the voxel $v$ can be removed only if for all $(x,y) \in v$, $F_{disk}(x,y) < 0$, or equivalently
\begin{equation}
    \max_{(x,y)\in v} F_{disk}(x,y)<0.
\end{equation} 
Although, in general, finding the maximum of $F(x,y)$ for $(x,y)$ inside a voxel analytically can be difficult, it is enough to use an upper limit of $F(x,y)$: $\tilde{F}(x,y) \ge F(x,y)$, as long as $\tilde{F}(x,y)$ tends to $F(x,y)$ when the voxel size goes to $0$. The Zhang algorithm used for generating saturated RSA packings of two-dimensional regular polygons \cite{Zhang2018} employs this approach. The following is its brief summary. 
Two polygons intersect only if any of their edges intersect. Assuming that there are two line segments: $A$ and $B$, each defined by two vectors pointing at their two ends, $(\bf{a_1}, \bf{a_2})$ and $(\bf{b_1}, \bf{b_2})$, respectively. Segments $A$ and $B$ intersect if and only if the ends of segment $A$ lay on the opposite sides of the line containing segment $B$, and vice versa. Thus,
\begin{widetext}
\begin{eqnarray}
    F_1\left( \bf{a_1} ,\bf{a_2} ,\bf{b_1}, \bf{b_2} \right)  = f\left( \bf{a_1} ,\bf{a_2} ,\bf{b_1} \right) \cdot f\left(\bf{a_1} ,\bf{a_2} ,\bf{b_2} \right)  & < &  0,  \mbox{ and}  \nonumber \\
    F_2\left( \bf{a_1} ,\bf{a_2} ,\bf{b_1}, \bf{b_2} \right)  = f\left(\bf{b_1} ,\bf{b_2} ,\bf{a_1}\right) \cdot f\left(\bf{b_1} ,\bf{b_2} ,\bf{a_2}\right)  & < &  0,
    \label{eq:intersection}
\end{eqnarray}
\end{widetext}
where $f\left( \bf{a} ,\bf{b} ,\bf{c} \right) = \left( \bf{c}-\bf{a} \right) \cdot \left( \bf{b}-\bf{a} \right)$ is positive if the point $\bf{c}$ is on one side of the line passing through $\bf{a}$ and $\bf{b}$, and negative if $\bf{c}$ is on the other side.

When generating two-dimensional RSA packings of polygons, the placement of an incoming particle can be specified by three scalars: the coordinate of its center, $(x, y)$, and the angle between the orientation of the polygon and a reference orientation, $\phi$. Thus, each point of a three-dimensional auxiliary space $(x, y, \phi)$ corresponds to an insertion of a trial polygon in this particular location and orientation. Three-dimensional voxels are employed to trace parts of this space where successful insertion trials are possible. In subsequent stages of the algorithm, some of these voxels are covered by newly added polygons; others are divided into $2^3$ subvoxels to approximate the available space better. As mentioned above, the crucial stage of the algorithm is determining if a voxel can be removed as inserting a new polygon inside any of its points leads to an intersection with previously placed objects.
The algorithm identifies polygon overlaps by checking if any sides of two polygons overlap using Eq.~(\ref{eq:intersection}). To determine if a given voxel overlaps with an existing polygon, the algorithm treats the voxel as a trial insertion with an uncertainty $(x\pm \delta x, y\pm \delta y, \phi\pm \delta \phi)$, and performs a worst-case error analysis of the intersection criterion. The analysis is based on the functions $\delta F_i\left( \bf{a_1} ,\bf{a_2} ,\bf{b_1}, \bf{b_2} \right)$, which provide the error bounds of conditions (\ref{eq:intersection}) for edge $A$ belonging to a polygon already added to the packing and edge $B$ from a trial particle within the voxel (see Eqs. (8)-(15) in ref. \cite{Zhang2018}). When $\tilde{F}_i = F_i + | \delta F_i| < 0$, an intersection always exists no matter how $\delta x$, $\delta y$, and $\delta \phi$ vary within their respective limits (determined by the voxel size); and thus the entire voxel can be removed. Additionally, if the condition (\ref{Fdisk}) is fulfilled for the existing polygon inscribed circle, a voxel was removed too. 

\begin{figure}[htb]
    \centering
    \includegraphics[width=0.7\columnwidth]{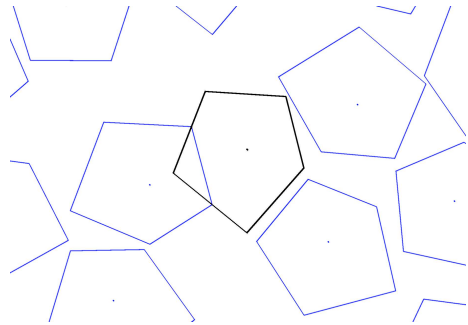}
    \caption{An unavailable voxel that could not be identified as such using the original algorithm \cite{Zhang2018}. Blue pentagons are objects already inserted into the packing and the black pentagon represents the unavailable voxel.}
    \label{fig:undeterminedVoxel}
\end{figure}

It turns out that for non-regular polygons, this algorithm requires a modification. If one employs the original algorithm, certain unavailable voxels can never be correctly identified.
\begin{figure}[htb]
    \vspace{0.2in}
    \centering
    \includegraphics[width=0.7\columnwidth]{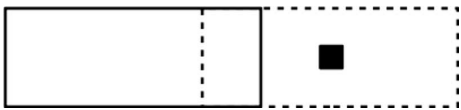}
    \caption{Example of a voxel (the black square) that would never be removed by the algorithm described in \cite{Zhang2018}. The solid rectangle is already part of the packing while the dashed one is the virtual one with the center inside the black square voxel. Rectangles partially overlap but their edges do not, according to the criterion (\ref{eq:intersection}).}
    \label{fig:counterexample}
\end{figure}
For example, the black voxel in Fig. \ref{fig:counterexample} will never be removed as it contains a point (corresponding to the center of the dashed rectangle) for which the functions $F_i = 0$. Therefore, even if $| \delta F_i| \to 0$ with decreasing voxel size, the maximum over the voxel is $\tilde{F}_i \ge 0$. Thus, the voxel will not be eliminated even though there is no possibility to place the non-intersecting rectangle in it.
A similar situation is shown in Fig.~\ref{fig:undeterminedVoxel}, where edges of one object cross vertices of the other one. While both objects intersects, their edges, according to (\ref{eq:intersection}), do not.
Again, a small voxel containing the center of the black polygon should be removed because the black polygon intersects with the existing blue polygon. However, the program could not ascertain its unavailability since the black polygon's sides only touch the sides of the blue polygon. In this case, the functions $F_i$ are again equal to zero, and their upper bound $F_i + \delta F_i \ge 0$. Thus, this voxel will not be eliminated and the algorithm will never stop. It is theoretically possible that such trial object will be randomly selected, which would lead to packing with overlapped objects. Practically, the probability of such selection is negligible $\sim (1/2^{64})^3$, assuming that three double-precision variables determine the position and orientation of the object. Even if weak inequalities in (\ref{eq:intersection}) are replaced by strong ones, which is the protection against overlapping of packed objects, the problem with voxels that can never be removed still remains. 

\begin{figure}[htb]
    \centering
    \includegraphics[width=0.7\columnwidth]{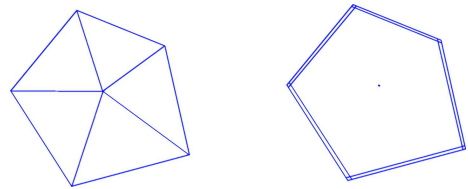}
    \caption{Illustration of the two ways to add helper segments to an irregular pentagon to solve the problem detailed in the text.}
    \label{fig:helperSegments}
\end{figure}

In this paper, we present a way to fully overcome this issue. We place so-called ``helper segments,'' which are additional line segments, within a polygon.  Although there might be many different ways of adding helper segments to overcome this problem, here we present two generic and numerically efficient methods. The first method is to add helper segments from an arbitrary point inside the polygon to all vertices. The second method is to add a parallel helper segment behind each of the polygon's edge. The distance between the edge and the helper segment parallel to it should be smaller than the distance between any two vertices. Our experience shows that the first method is generally faster, since the helper segments share one endpoint with polygon edges, and therefore do not need not be recalculated. However, the second method may be required when the first method is inapplicable. (For example, if it is impossible to find a point inside a particular non-convex polygon such that all helper segments from the point to its vertices are inside the polygon). The two ways to add helper segments are illustrated in Fig.~\ref{fig:helperSegments}.

Another possibility is to treat all polygon diagonals as helper segments, which works well for rectangles. However, the number of diagonals grows quadratically with the number of vertices, which significantly reduces the efficiency of the algorithm for polygons with  high number of vertices.  
\section{Test case - RSA of triangles}
To prove the correctness of the presented algorithm, saturated packings built of identical triangles were generated in order to find a shape that provides the densest packings.

\subsection{Model}

\begin{figure}[htb]
    \centering
    \includegraphics[width=0.7\columnwidth]{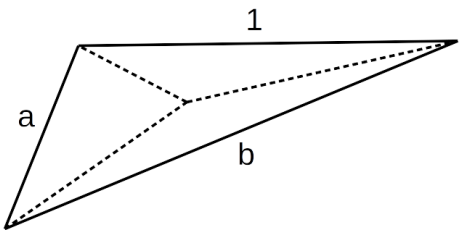}
    \caption{Model of a triangle of side lengths ratio $a<1<b$. Dashed lines are helper segments from the triangle mass center to its vertices.}
    \label{fig:triangle-model}
\end{figure}

The shape of an arbitrary triangle is fully determined by three numbers corresponding to its side sizes. Without any loss of generality, the length of the middle side can be taken as the unit of length. Thus, there are only two independent variables $a$ and $b$ such that $ a \le 1 \le b$, and additionally, due to triangle inequality condition, $a + 1 \ge b$ (see Fig. \ref{fig:triangle-model}).
Triangles were rescaled to have a unit surface area, and were put randomly on a square box of size $S=300\times 300$. Periodic boundary conditions were used as it has been shown that they reduce finite-size effects \cite{Ciesla2018}. For each shape of triangle, $N=100$ independent, saturated packings were generated. The mean packing fraction was equal to $\theta = (1/N)(\sum_{i=1}^N n_i/S)$, where $n_i$ was the number of triangles in $i-$th packing. The statistical error of $\theta$, given by $\sigma(\theta) = (1/N)\left[\sum_{i=1}^N (\theta - n_i/S)^2\right]^{1/2}$, was below $10^{-4}$. During packing generation, three helper segments form the mass center of the triangle to its vertices were used -- see Fig. \ref{fig:triangle-model}.

\subsection{Packing fractions}

\begin{figure}[htb]
    \centering
    \subfigure[]{\includegraphics[width=0.3\columnwidth]{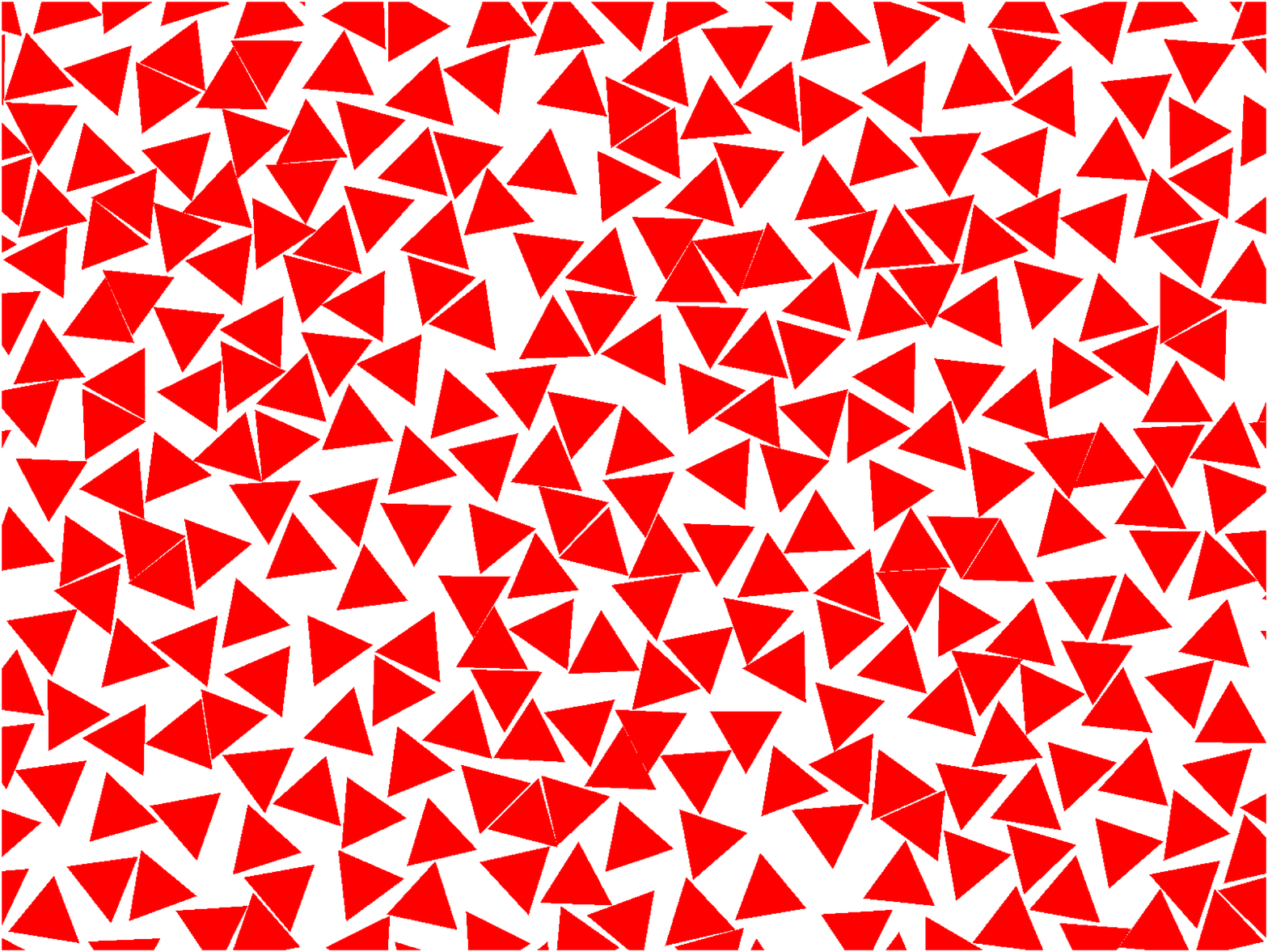}}
   \subfigure[]{\includegraphics[width=0.3\columnwidth]{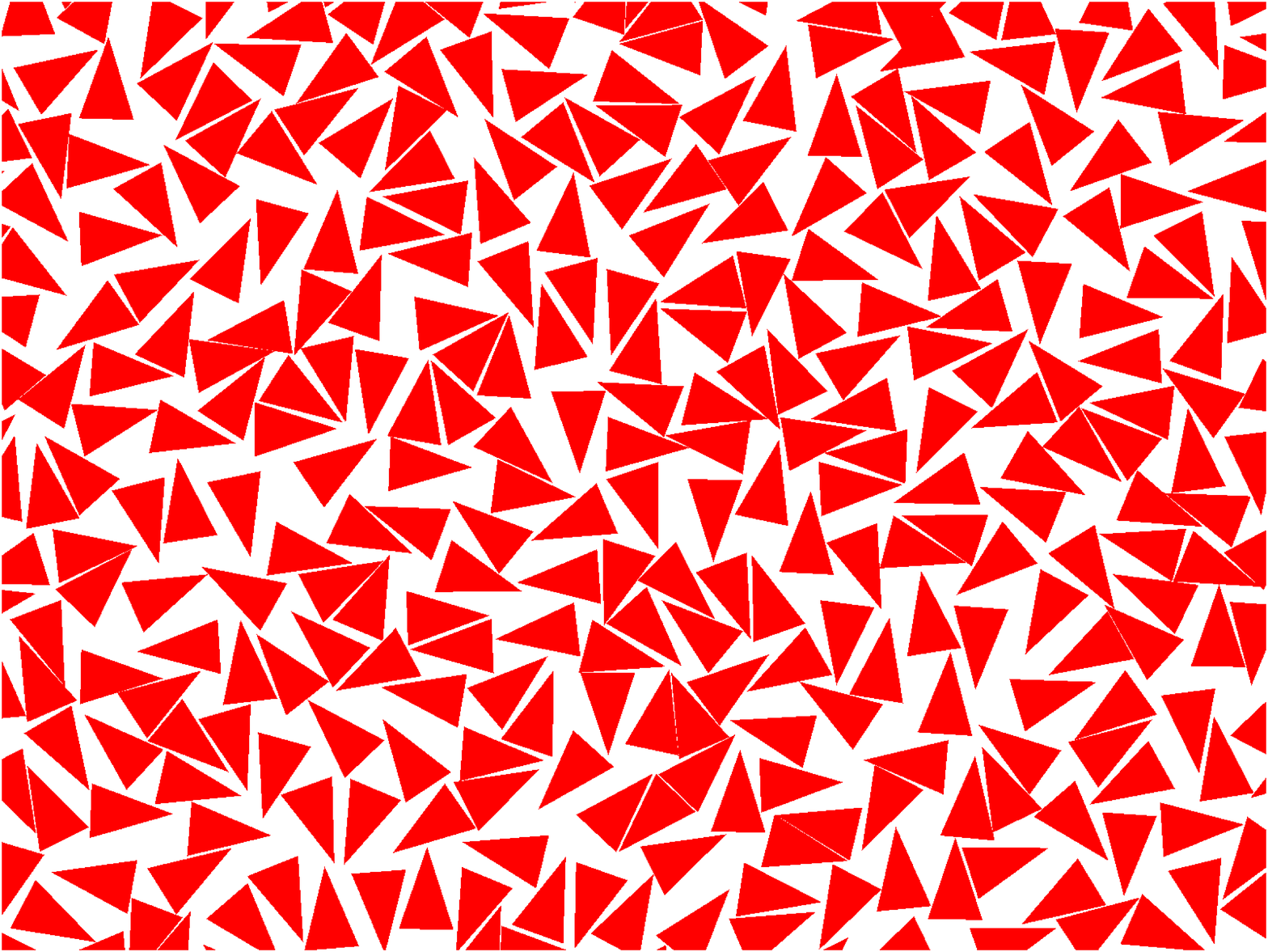}}
   \subfigure[]{\includegraphics[width=0.3\columnwidth]{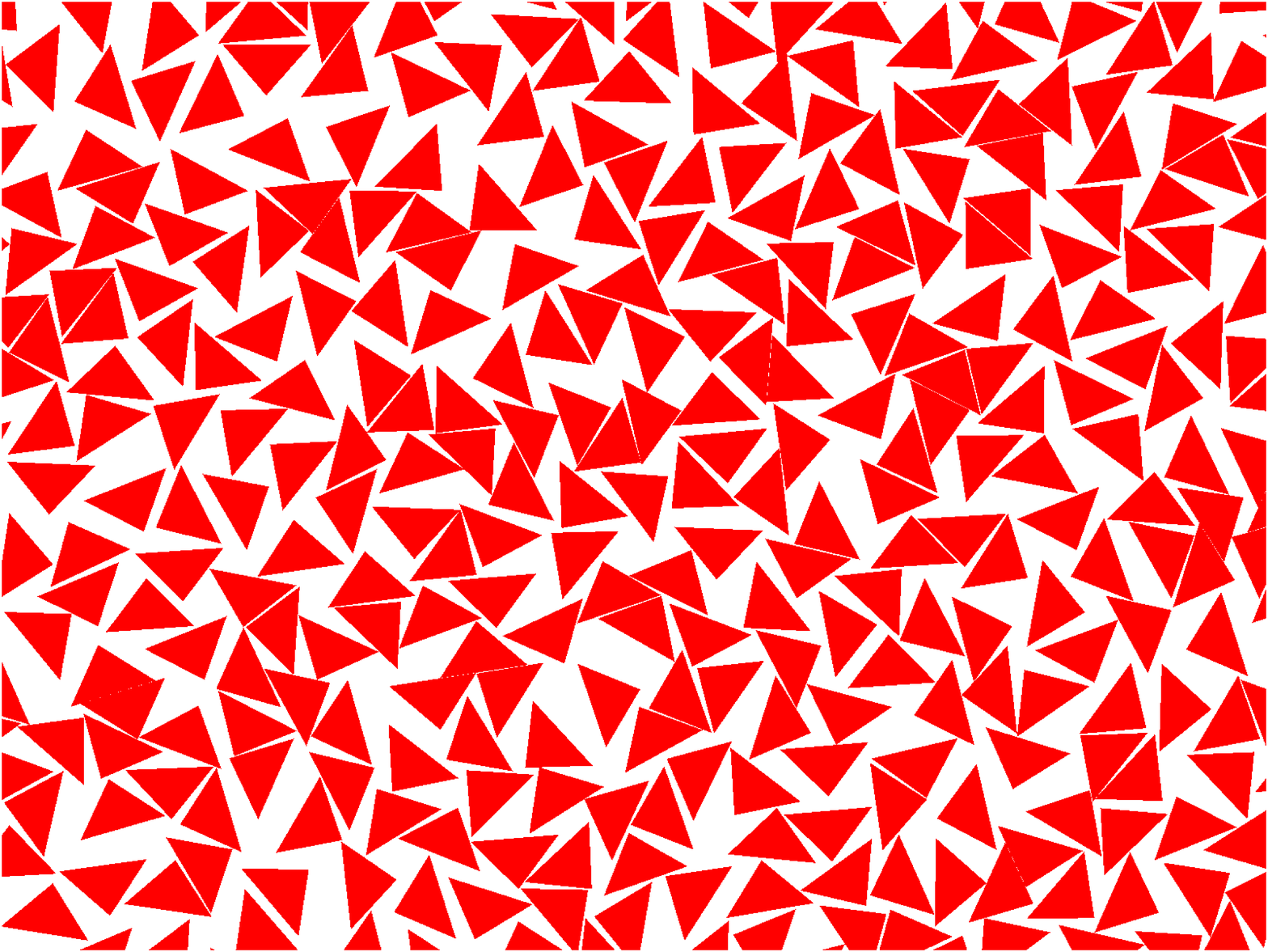}}
    \caption{Fragments of example saturated random packings of rectangles for (a) $a=b=1.0$; (b) $a=0.63$, $b=1.0$; and (c) $a=1.0$, $b=1.31$.}
    \label{fig:examplepackings}
\end{figure}

\begin{figure}[htb]
    \centering
    \includegraphics[width=0.7\columnwidth]{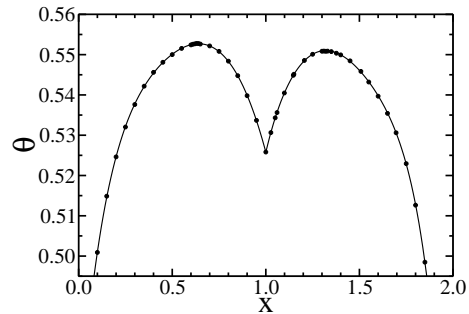}
    \caption{The mean saturated packing fraction for isosceles triangles. The parameter $x$ denotes the length ratio of isosceles triangle base to its arm, e.g., $x=1$ corresponds to equilateral triangle. Dots are numerical results, and the solid line is to guide the eye. Statistical errors are smaller than the size of dots.}
    \label{fig:isosceles}
\end{figure}

Fragments of example packing are shown in Fig.\ref{fig:examplepackings}.
The analysis of the obtained values of the mean saturated packing fractions started with the case of isosceles triangles of side lengths $1, 1, x$, where $x$ can be any positive value. The results are shown in Fig. \ref{fig:isosceles}.
The highest packing fraction is reached for $x=0.63$ and equals $\theta = 0.552814 \pm 0.000063$. Slightly lower value is observed for $x=1.31$, were $\theta=0.550906 \pm 0.000057$. The local minimum is present for the equilateral triangle, which agrees with theoretical arguments that shapes of higher symmetry form less dense packings \cite{Baule2013}. Similar behavior has been observed for other anisotropic shapes \cite{Vigil1989, Viot1992, Haiduk2018, Kasperek2018}. The packing fraction for equilateral triangle is $\theta= 0.525818 \pm  0.000059$ which agrees with previously reported values \cite{Ciesla2014, Zhang2018}.

\begin{figure}[htb]
    \centering
    \includegraphics[width=0.7\columnwidth]{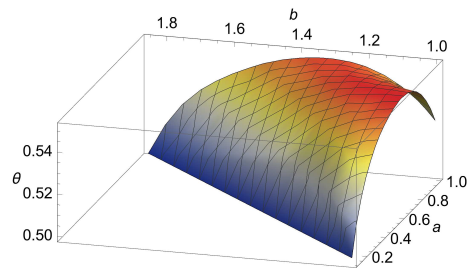}
    \caption{The mean saturated packing fraction for arbitrary triangles.}
    \label{fig:packingfraction}
\end{figure}

The results for packings built of arbitrary triangles are shown in Fig.~\ref{fig:packingfraction}.
Interestingly, for all studied sets of parameters $(a,b)$ not a single packing was found that was denser than the one obtained for isosceles triangles. Besides, the maximum in the $(a,b)$ plane seems to be almost flat along the direction that connects two maxima for isosceles triangles. Therefore, at this level of accuracy, it is hard to determine if the global maximum is reached for the isosceles triangle or for another shape that is close to $(a=0.63, b=1)$. However, it is unexpected that shape of the higher symmetry does not correspond to a local minimum of the packing fraction \cite{Baule2013}. 
\subsection{Microstructural properties}

\begin{figure}[htb]
	\centering
    \vspace{0.5in}
	\includegraphics[width=0.7\columnwidth]{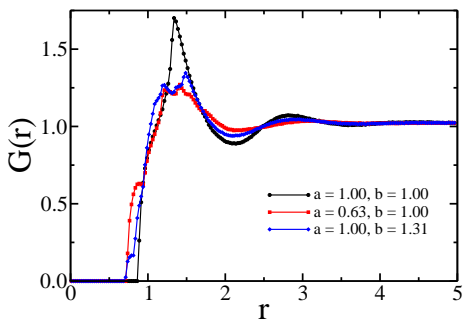}
    \caption{Density autocorrelation functions for packings built of three different triangles.}
    \label{fig:correlations}
\end{figure}

Microstructural properties of the obtained packings can be studied using density autocorrelation function:
\begin{equation}
    G(r) = \lim_{dr \to 0} \frac{N(r, r+dr)}{2\pi r \, dr \, \theta},
\end{equation}
where $N(r, r+dr)$ is the mean number of triangles with the mass center at the distance between $r$ and $r+dr$ from the center of a reference particle. The results for sample packings are shown in Fig. \ref{fig:correlations}.
The behavior observed is typical for random media. For equilateral triangle the first maximum is higher and sharper due to the symmetry -- when two triangles are touching their sides, the distance between their centers does not depend on which sides are in touch. For isosceles triangles, this is not true, and therefore instead of one maximum, two smaller maxima are observed. For larger distances, $G(r)$ oscillations are damped. It has been shown analytically that for one dimensional RSA packings these oscillations vanish super-exponentially with the distance \cite{Bonnier1994}. Here, $G(r) \approx 1$ for $r>5$. 

\subsection{Finite-size effects}
\begin{figure}[htb]
	\centering
    \vspace{0.5in}
	\includegraphics[width=0.7\columnwidth]{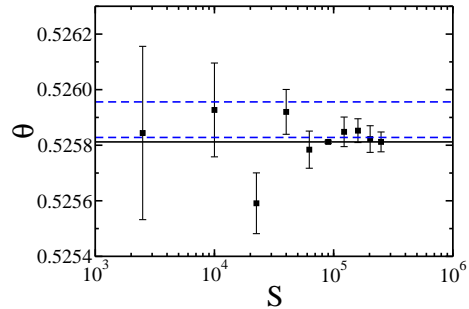}
    \caption{The dependence of saturated mean packing fraction on packing size. For all packing sizes except $S=90000$, packing fraction was calculated by averaging over $100$ independent random packings. For $S=90000$, $10^4$ independent packings were used. Error bars correspond to the standard deviation of the mean packing fraction. Solid black line corresponds to $\theta(S=9000) = 0.5258119$. Dashed blue lines correspond to the range of $\theta$ determined by Zhang \cite{Zhang2018}.}
    \label{fig:q_size}
\end{figure}

It has been shown for RSA of disks that the measured value of saturated packing fraction for small packings oscillates with respect to the system size, similar to the oscillations found in the tail of the autocorrelation function \cite{Ciesla2018}. Therefore, it was expected (as a packing side length used in the study is much larger than $5$) that the finite-size effects should not affect obtained values of packing fraction. To prove that, the dependence of packing fraction on packing size for equilateral triangles was analyzed. The results are shown in Fig. \ref{fig:q_size}.
The result from our previously-used system size, $S=90000$ certainly agrees with the best estimation given by the solid black line. Here we also present the packing fraction range reported earlier in \cite{Zhang2018}, which is slightly higher but very close to the one reported here. This can happen with approximately one-third probability due to the probabilistic interpretation of standard deviations.

\subsection{Kinetics of packing growth}
\begin{figure}[htb]
	\centering
    \vspace{0.5in}
	\includegraphics[width=0.7\columnwidth]{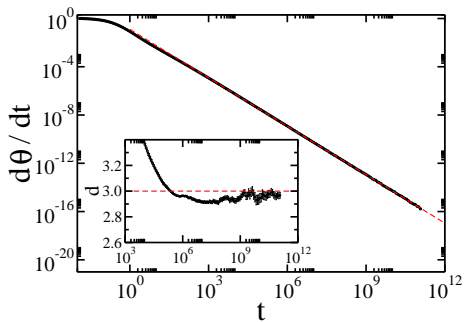}
    \caption{The dependence of increments of packing fraction on dimensionless time for equilateral triangles. Black dots are measured data and red dashed line is a power-law fit $d\theta / dt = 0.12728 \, t^{-1.3387}$. Inset shows the value of parameter $d$ estimated from data obtained for times within $[0.01t, t]$. Red dashed line corresponds to $d=3$.}
    \label{fig:dq_t}
\end{figure}

Kinetics of RSA packing growth near saturation depends on the number of iterations according to the power-law \cite{Pomeau1980, Swendsen1981}
\begin{equation}
  \theta(t) = \theta - A t^{-\frac{1}{d}}.
  \label{eq:fl}
\end{equation}
The number of iterations is often measured in the so-called dimensionless time, which unit corresponds to $S/S_p$ iterations, where $S_p$ is the area covered by a single particle in the packing. Such scaling does not affect the exponent in the above power-law and is commonly used to compare RSA kinetics of differently sized packings. 
Although the parameter $d$ in (\ref{eq:fl}) is simply the space dimension for spheres, Baule has shown recently that when anisotropic particles are placed according to the RSA protocol in such a way that their centers lay on a one-dimensional line, $d$ depends on properties of contact function of packed shapes. Parameter $d$ is noticeably bigger when this function is not analytic \cite{Baule2017}. For two-dimensional non-spherical RSA packings, there is numerical evidence that $d=3$ for RSA of ellipses \cite{Viot1992, Ciesla2016}, or rectangles \cite{Viot1990, Vigil1990}, however, these studies do not reach saturation limit. In this study, using the algorithm described above, we can check the validity of (\ref{eq:fl}) and calculate $d$ arbitrarily close to saturation. For all examined shapes, the determined value of parameter $d$ was in the range of $(2.79, 3.17)$ with the fitting error at the level of $0.05$. Fig. \ref{fig:dq_t} shows example data obtained for saturated packings built of equilateral triangles. 
The slope of the dependence is almost constant (see inset in Fig. \ref{fig:dq_t}) for large enough $t$, and $d \approx 3$.

\begin{figure}[htb]
	\centering
    \vspace{0.5in}
    \includegraphics[width=0.7\columnwidth]{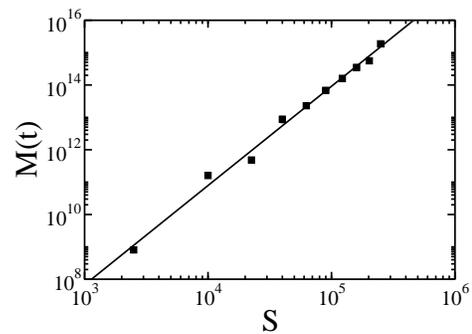}
    \caption{The dependence of the median of the number of iterations (expressed in dimensionless time) needed to generate saturated RSA packing built of equilateral triangles on packing size. Squares represent numerical data obtained from $100$ packings. Solid line is a power fit: $M(t) = 0.043177 \, S^{3.0649}$.}
    \label{fig:median}
\end{figure}

For saturated packing of spherically symmetric particles, it was previously established that parameter $d$ can also be obtained from the dependence of the median of iterations needed to generate saturated packings expressed in dimensionless time units on packing size \cite{Ciesla2017}:
\begin{equation}
    M(t) \sim S^d.
\end{equation}
As shown in Fig. \ref{fig:median}, the above relation is also valid for equilateral triangles, and, because its derivation bases on very general assumptions \cite{Ciesla2017}, we expect it to be valid for other anisotropic shapes.

Figure \ref{fig:median} also confirms that $d$ for triangles is the same as for other anisotropic shapes with analytic contact function. Thus, the result obtained by Baule for RSA on a one-dimensional line \cite{Baule2017} cannot be directly extended to higher dimensions.

\begin{figure}[htb]
	\centering
    \vspace{0.5in}
	\includegraphics[width=0.7\columnwidth]{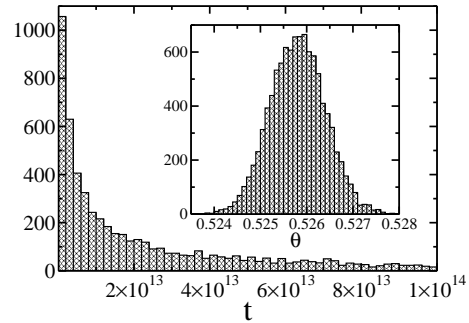}
    \caption{Histograms of the number of iterations (expressed in dimensionless time) needed to generate saturated packing and the packing fraction (inset) obtained from studying $10^4$ packings built of equilateral triangles.}
    \label{fig:histograms}
\end{figure}

At last, we study the distributions of packing fractions and the number of iterations needed to generate saturated packings. They are important because the full distribution provides extra information beside the mean and standard deviation. Here, as we can see in Fig. \ref{fig:histograms}, the packing fraction is normally distributed, while a power-law gives the distribution of the number of iteration needed to saturate packing. 
It is noteworthy that the same was observed for RSA of disks \cite{CieslaNowak2016, Ciesla2017}. Thus again, this seems to be a universal property for RSA of arbitrary shapes.
\section{Summary}
This study presents an algorithm that allows generating saturated random packings according to the RSA protocol for a wide range of shapes. The algorithm is based on the idea recently introduced in \cite{Zhang2018}, and incorporates an additional step of adding helper segments to a polygon in order to determine if another trial particle can be added to the packing in a given area. 

This algorithm was used to generate saturated random packings of triangles to find the shape that yields the densest packings. It appeared that the highest packing fraction, $0.552814 \pm 0.000063$, was observed for isosceles triangles of side lengths ratios $0.63:1:1$. Also, a previous estimation of the mean packing fraction for equilateral triangles \cite{Zhang2018} is improved to $0.5258119 \pm 0.0000059$.

The presented algorithm was also used to check kinetics of the growth of RSA packing built of triangles near saturation, and it was found that in contrast to one-dimensional RSA, it is governed by the same exponent as for other smooth shapes. Additionally, it has been shown that the number of iterations needed to saturate packings scales with packing size according to a power-law -- the same as for spherically symmetric particles. 
\section*{Acknowledgments}
MC and PK acknowledge the support of grant no. 2016/23/B/ST3/01145 of the National Science Center, Poland. A part of numerical simulations was carried out with the support of the Interdisciplinary Center for Mathematical and Computational Modeling (ICM) at the University of Warsaw under grant no.\ GB-76-1.
GZ thanks the U.S. Department of Energy, Office of Basic Energy Sciences, Division of Materials Sciences and Engineering under Award DE-FG02-05ER46199.
\section*{Supplementary materials}
%
%The Mathematica notebook attached contains all measured mean packing fractions together with their standard deviations. 
The source code of the program generating saturated random packings of arbitrary polygons and in particular triangles, is available on \url{https://github.com/misiekc/polygonRSA}.
%
%\section*{References}
\bibliography{main}
\end{document}